\documentclass[aps,pre,preprint]{revtex4-1}
\usepackage{amsmath}
\usepackage{amsfonts}
\usepackage{amssymb}
\usepackage{graphicx}
\usepackage{color}
\usepackage{ulem}
\usepackage{soul}

\newcommand{\prs}[1]{{\left(#1\right)}}
\newcommand{\col}[1]{{\left[#1\right]}}
\newcommand{\chs}[1]{{\left\{#1\right\}}}
\newcommand{\avg}[2]{{\left<#1\right>_{#2}}}

\newcommand{\sech}{{\mbox{sech}}}

\newcommand{\sgn}{{\mbox{sgn}}}

\newcommand{\tbs}{{\tilde{\beta}_\sigma}}
\newcommand{\tbt}{{\tilde{\beta}_\tau}}
\newcommand{\bes}{{\beta_\sigma}}
\newcommand{\bet}{{\beta_\tau}}
\newcommand{\tJ}{{\tilde{J}}}
\newcommand{\cut}[1]{{}}
\newcommand{\new}[1]{\color{blue}{}}
\renewcommand{\emph}{}
\begin{document}

\title{Interacting Non-equilibrium Systems with Two Temperatures}
\author{Roberto C. Alamino, Amit Chattopadhyay and David Saad}
\affiliation{Non-linearity and Complexity Research Group, Aston University, Birmingham B4 7ET, UK}

\begin{abstract}
We investigate a simplified model of two  fully connected magnetic systems maintained at different temperatures by virtue of being 
connected to two independent thermal baths while simultaneously being inter-connected with each other. Using generating functional 
analysis, commonly used in statistical mechanics,
we find exactly soluble expressions for their individual magnetisations that define a two-dimensional non-linear map, the
equations of which have the same form as those obtained for densely connected equilibrium systems.
Steady states correspond to the fixed points of this map, separating the parameter space into a rich set of
non-equilibrium phases that we analyse in asymptotically high and low (non-equilibrium) temperature limits. The theoretical formalism 
is shown to subvert to the classical non-equilibrium steady state problem for two interacting systems with a non-zero heat transfer 
between them that catalyses a phase transition between ambient non-equilibrium states.
\end{abstract}

\maketitle

\section{Introduction}
\label{section:Intro}

The study of energy transport between interacting non-equilibrium systems, each in contact with heat baths maintained at different 
temperatures, has a long history~\cite{Politi03, Dhar08, Lebowitz08,Dhar11,Kundu11} spanning a multitude of non-equilibrium systems. 
Two macroscopic systems in thermal contact with each other will exchange energy until equilibrium is reached as long as they are kept 
thermally isolated from the environment. However, if systems are coupled to independent thermal reservoirs that are maintained at two 
different temperatures, they can be kept indefinitely in a non-equilibrium steady state (NESS) without any perceptible change in 
macroscopic \emph{local} properties. While the definition of a `non-equilibrium temperature' has remained an open question for long, not the 
least due to the absence of an inherent Hamiltonian description, generalised manifestations of the fluctuation-dissipation 
theorem~\cite{Kawasaki67,Bochkov81,Jarzynski97,Chattopadhyay11} have often addressed the issue of non-zero thermal flux across 
non-equilibrium quasistatic systems that are driven by external 
time-dependent forces. The common topic of interest in all these studies has been the depiction of the temperature profile as a 
function of the free energy differences of connected systems and the dependence of associated heat flows on this `non-equilibrium 
temperature'.

A recent series of studies have specifically focused on stochastic fluctuations~\cite{Dhar08,Lebowitz08,Dhar11} as the initiators of 
external forcing that are simultaneously capable of maintaining the non-equilibrium energy traffic across the connected (non-equilibrium) 
systems, both in linear~\cite{Dhar08,Kundu11} and non-linear regimes~\cite{Saito11}. The models considered include harmonic chains (in one, 
two and three dimensions) and harmonic crystals with (mostly) white-noise Langevin reservoirs connected at the ends of the chains/crystals. 
All these works indicate an analytical dependence of the non-equilibrium heat transfer on the dimensionality of the system, categorically 
characterising the phase space of the combined system into either of the three standard universality classes - ballistic, diffusive and 
anomalous (sub or super-diffusive). The non-equilibrium manifolds inclusive in  these studies are also known to abide the 
celebrated `additivity principle'~\cite{Derrida04}.

Interestingly enough, none of these studies considered externally forced spin systems as the perturbed basis, be it in the form of the 
paradigmatic spin-1/2 Heisenberg system, or its equivalent spherical $n$-spin model. In view of the wide array of spin and spin-glass 
type systems undergoing non-equilibrium energy exchanges that lead to steady state that exhibits metal-insulator-like conduction, this 
appears a unique exclusion. One must mention~\cite{Sen08} as an exception in this regard; where starting from spinless quantum dots, the 
authors were able to prove the existence of two-particle scattering states leading to a steady non-zero Landauer current. While of a 
different flavour, another work worth mentioning is the interpretation of the replica method as a two temperature system that exhibits 
different dynamics related to the corresponding effective temperatures of annealed and quenched 
variables~\cite{Dotsenko93,Dotsenko94,vanMourik01}.

Theoretical implications aside, the two-temperature formalism we examine here has a multitude of direct applications in condensed matter 
systems. Many solids, including the high-$T_c$ cuprates and pnictides superconductors, have layered structures where intra- and inter-layer 
couplings are different, leading to the existence of two different temperature baths in an inter-connected spin system~\cite{Mourachkine02}. 
As would be detailed later, a corollary of our model addresses the simplified case of a temperature gradient perpendicular to the layers 
that effectively embeds anharmonicity in the interacting spin chains. Identically, superlattices subjected to anharmonic (and often aperiodic) 
forcing while being maintained at two different temperatures, also fall under the more general purview of the model presented 
here~\cite{Schuller80}. Interestingly, other potential applications
in the study of dark matter, where it has been suggested~\cite{Spergel00} that normal and dark matter are coupled to each other through 
gravity while remaining in contact to thermal reservoirs at different temperatures in their respective sectors. Some~\cite{Feng08} even 
propose that the actual equilibrium temperature of each sector can be different.

As a major departure from the existing trend, in this article our NESS connected systems are {\it deterministically forced} and time 
independent to start with. More specifically, we consider two mutually connected spin-1/2 chains that are individually attached to two 
separate heat reservoirs that are maintained at two separate fixed temperatures through steady supplies of external energy. Our interest 
here is in the magnetisation regime of each system under the action of mutual (heat) flux exchange and how such non-equilibrium magnetisation 
connects to equilibrium steady states, a direct allusion to the alternative fluctuation-dissipation formalism.

The experimental analogue of our theoretical framework is that of two inter-connected magnetic systems adiabatically maintained at two 
different constant temperatures by virtue of being connected to independent temperature baths. The two magnetic systems that are suffused 
with non-zero magnetisations at the initial time instant $t=0$ are then allowed to interact via magnetic couplings and left to evolve in 
time. Physically, this
can be accomplished by bringing two such magnetic systems close together in vacuum or by connecting them through some thermal insulating
material. The steady states of this composite system will then be analysed under synchronous discrete Markovian dynamics.

The microscopic model and its dynamical rule will be presented in section~\ref{section:DM}, the solution of which will be a coupled
two-dimensional dynamical map for the magnetisations of the two subsystems. As we are interested mainly
in the steady state solutions, we will analyse in some depth the fixed points of this map on section~\ref{section:FP}. 
Section~\ref{section:PP} contains a stability analysis of the global paramagnetic phase of the model.
Further discussion of the results, their relevance and future directions are presented in section~\ref{section:Conc}.

\section{The Dynamical Model}
\label{section:DM}

We start by considering a two-dimensional fully connected Heisenberg lattice where each node represents a spin-1/2 particle. We then identify 
two different sets of nodes, $\mathbf{\sigma}$ and $\mathbf{\tau}$, which we refer to as the $\sigma$ and $\tau$ \emph{components} of the 
system (subsystems). We ascribe the rule that nodes in each one of the components are fully connected to their own separate thermal 
reservoirs and no particle is connected to more than one particle of the other subsystem. 
The reservoirs to which the
$\sigma$ and $\tau$ components are coupled to temperatures $T_\sigma$ and $T_\tau$, respectively. We assume that the $\sigma$-component 
has $N_{\sigma}$ nodes $\sigma_i\in\chs{\pm1}$ and, similarly, the $\tau$-component has $N_{\tau}$ nodes
$\tau_i\in\chs{\pm1}$.

The system dynamics will be taken as discrete and defined by the local stochastic update rules for each individual
spin-node by the probabilities
\begin{align}
  P[\sigma_i(t+1)] &\propto \exp\chs{\bes \sigma_i(t+1)h^\sigma_i[\sigma(t),\tau(t)]},\\
  P[\tau_i(t+1)]   &\propto \exp\chs{\bet \tau_i(t+1)h^\tau_i[\sigma(t),\tau(t)]},
\end{align}
with $\beta_a=1/T_a$,  $a=\sigma,\tau$, where we are using units such that the Boltzmann constant is $k_B=1$. The local fields
acting on each spin are defined as
\begin{align}
  h^{\sigma}_i &= \frac{J_\sigma}{N_\sigma} \sum_{j\neq i} \sigma_j+J'_{\sigma\tau}\tau_i,\\
  h^{\tau}_i   &= \frac{J_\tau}{N_\tau} \sum_{j\neq i} \tau_j+J'_{\tau\sigma}\sigma_i~.
\end{align}
The above description suggests that constituents of each component interact with one another via long-range (mean-field) interactions $J_a$ 
and with spins from the other component locally via the short-ranged (local) interactions $J'_{ab}$. For simplicity, we assume that 
$J_\sigma=J_\tau=J$, $J'_{\sigma\tau}=J'_{\tau\sigma}=J'$ and that the number of spins in both components is the same $N_\sigma=N_\tau=N$.

\begin{figure}
  \centering
  \includegraphics[width=10cm]{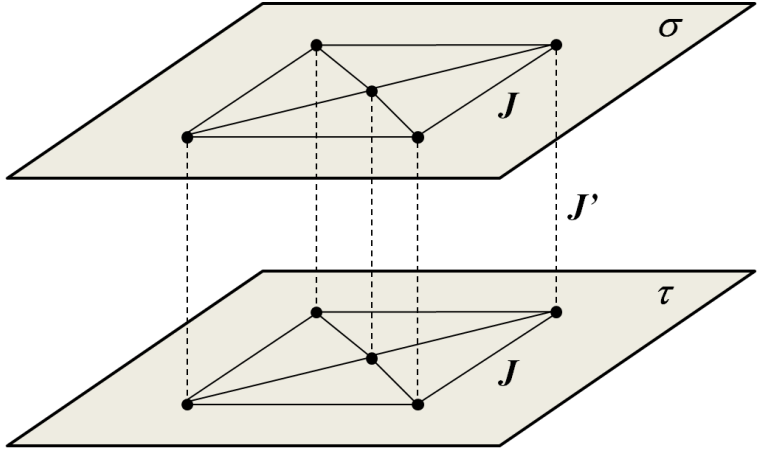}
  \caption{Visualisation of the model as two planes $\sigma$ and $\tau$. The long-ranged intraplane interaction is $J$ and the
           short-ranged interplane interaction is $J'$.}
  \label{figure:DM}
\end{figure}

A helpful way of visualising this system is depicted in Fig.~\ref{figure:DM} where we consider the $\mathbf{\sigma}$ and $\mathbf{\tau}$
components lying on two different parallel planes. We will use this graphic interpretation as it conveniently separates the global and 
local interactions and simplifies the nomenclature we will be using. The long-range interaction defined by the coupling $J$ will be 
termed \emph{intraplane interaction}, while $J'$ will define the local and symmetric \emph{interplane interaction}.

The system is updated synchronously, meaning that at each time step all the spins are updated simultaneously;
the generating functional method~\cite{Mozeika12} then allows for the dynamics to be solved exactly. This results in a two-dimensional 
non-linear dynamical map for the magnetisations of the different components of the system given by
\begin{align}
  m^{\sigma}(t+1) &= \avg{\tanh \bes h^\sigma(t)}{\tau(t)},
  \label{equation:ms}\\
  m^{\tau}(t+1)   &= \avg{\tanh \bet h^\tau(t)}{\sigma(t)},
  \label{equation:mt}
\end{align}
where
\begin{align}
  h^{\sigma}(t) &= Jm^\sigma(t)+J'\tau(t),\\
  h^{\tau}(t)   &= Jm^\tau(t)+J'\sigma(t),
\end{align}
and where averages $\langle\bullet\rangle$ are with respect to the corresponding probability distributions
\begin{align}
  P[\sigma(t)]&= \frac12 \col{1+\sigma(t)m^\sigma(t)},\\
  P[\tau(t)]  &= \frac12 \col{1+\tau(t)m^\tau(t)}.
\end{align}

The steady state solutions are given by the fixed points of this map
\begin{align}
  m^{\sigma} &= \avg{\tanh \bes (Jm^\sigma+J'\tau)}{\tau},\\
  m^{\tau}   &= \avg{\tanh \bet (Jm^\tau+J'\sigma)}{\sigma},
\end{align}
which are exactly the same equations that describe, in equilibrium, the magnetisation of a spins system with long-range interactions subjected
to a random magnetic field. These fixed points are analysed in the following section.

\section{Steady States}
\label{section:FP}

Steady state solutions of the dynamics of our system will be given by the
fixed points of the two-dimensional map defined by equations (\ref{equation:ms}) and (\ref{equation:mt}). A notational
simplification of those equations can be achieved by using the reduced variables $\tilde{\beta}_a = J'\beta_a$ and
$\tJ=J/J'$, resulting in the following map
\begin{align}
  f(m^\sigma,m^\tau) &= (f^\sigma,f^\tau),  \nonumber\\
  f^\sigma           &= \frac12 \sum_s \prs{1+sm^\tau} \tanh \tbs\prs{\tJ m^\sigma+s}, \label{eq:fs_ms}\\
  f^\tau             &= \frac12 \sum_s \prs{1+sm^\sigma} \tanh \tbt\prs{\tJ m^\tau+s}, \nonumber
\end{align}
with $s\in\chs{\pm1}$. The dynamics now depends only on three parameters $\tbs$,
$\tbt$ and $\tJ$ and the initial conditions on the magnetisations. Note, though, that $\tbs$ and $\tbt$ may assume negative values depending 
on the sign of the interaction $J'$ and cannot be interpreted straightforwardly as simply rescaled inverse
temperatures.

It is possible to obtain the fixed points of these equations numerically for any parameter set.
Still, it is interesting to study the behaviour of the system in the limits of high and low temperatures, which can be
obtained analytically and exposes a very rich phase diagram. The latter represents different types of interactions between the systems 
components and provides insights into the corresponding macroscopic properties.

The system, as defined, is symmetric under exchange of component labels $\sigma\leftrightarrow\tau$. Therefore, without loss of generality,
we choose the $\tau$-component to be connected to a thermal reservoir at
high temperature $\tbt\rightarrow0$. Then
\begin{equation}
 f^\tau \approx \tbt\prs{\tJ m^\tau+m^\sigma},
\end{equation}
resulting in the fixed point
\begin{equation}
\label{eq:mt_ms}
  m^\tau = \tbt\, m^\sigma,
\end{equation}
at leading order in $\tbt$.

Naturally, at high temperature the intraplane interaction $\tJ$ is not sufficiently strong
to influence the spins alignment and disappears from the equation. The interplane interaction still survives, although weakened by the 
thermal disorder, linking the magnetisations of the two components via a proportionality constant $\tbt$; the relative sign between the 
two is determined by the nature of the interaction $J'$ (ferromagnetic or antiferromagnetic) in an obvious way.

When the $\tau$-component temperature becomes infinite, with $\tbt$ being exactly zero, the intraplane interaction becomes
too weak to allow for the interaction with the $\sigma$-component to affect the $\tau$-component alignment, which then leads to a paramagnetic 
phase with magnetisation $m^\tau=0$. It is easy to see that consequently the $\sigma$-component becomes equivalent to a
mean-field Ising model in a random field of zero mean~\cite{nishimori01}.

Keeping the $\tau$-component's temperature fixed at a high but finite value, one can re-examine the system's properties at various 
temperature limits of the $\sigma$-component. For the case of high temperature (low $\tbs$), one obtains a proportionality relation between 
the magnetisations
\begin{equation}
  m^\sigma = \tbs \, m^\tau\Rightarrow  m^\tau=\tbt\tbs m^\tau.
\end{equation}

As we are assuming both temperatures to be high (both $\tilde{\beta}$'s small), the only solution is that both
magnetisations vanish at high enough finite temperatures. This signals a transition to
what we conventionally refer to as the \emph{global paramagnetic (GP) phase}. The GP phase is a ubiquitous fixed point,
present for \emph{any} temperature, but its uniqueness and stability are temperature-dependent.
In this section we will focus on its uniqueness property, i.e., analyse the existence of other possible
phases in these temperature limits. The issue of its stability will be addressed mainly in the next section.

\begin{figure}
 \center
 \includegraphics[width=10cm]{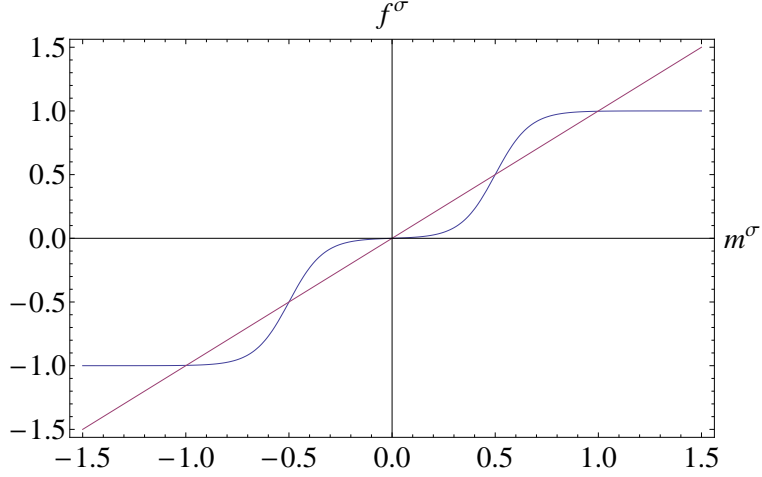}
 \caption{Fixed points of the magnetisation $m^\sigma$ for $J=2$ and $\tbs=3$}
 \label{figure:GPP}
\end{figure}

One notable feature of the high-temperature regime in $\tau$ is that it is rich enough to develop a total of either three
or five fixed points as the $\sigma$-component's temperature decreases, depending on the values of the intraplane
interaction $J'$. This can be obtained from the relation between $f_{\sigma}$ and $m_{\sigma}$ of Eq.~(\ref{eq:fs_ms}) by utilising the high 
temperature relation~(\ref{eq:mt_ms})and the intersection of $f_{\sigma}$ with the line $m_{\sigma}$. Figure~\ref{figure:GPP} shows an 
exemplar plot for $\tJ=2$ and $\tbs=3$. The graph shows five fixed points: a stable solution at $m^\sigma=0$, two stable solutions at high 
magnetisation values (largest) and two unstable solutions at intermediate values.

For $\tau$ at high temperature and at the limit $|\tbs|\rightarrow\infty$ one obtains
\begin{equation}
  m^\sigma = \avg{\sgn \, \tbs\prs{\tJ m^\sigma+\tau}}{\tau}=\sgn\,J'\cdot\avg{\sgn\prs{\tJ m^\sigma+\tau}}{\tau}.
\end{equation}

In this case, one has to consider the various possible values for the intraplane interaction $\tJ$. For $|\tJ|<1$
\begin{equation}
  m^\sigma = m^\tau\,\sgn\,\tbs = m^\tau\,\sgn\,J',
\end{equation}
which, combined with the proportionality between $m^\tau$ and $m^\sigma$, gives rise to a single solution $m^\tau=m^\sigma=0$ indicating
that the \emph{intraplane} interaction is too small to align the spins of the $\sigma$-system.
A straightforward, but lengthy consideration of the possibilities for $|\tJ|=1$ gives the same result. This means that there
are situations where the GP phase is the only phase present at the high-temperature $\bet \rightarrow 0$ limit for \emph{both} low and 
high $T_\sigma$; which reveals the importance of the relative (non-dimenionalised) interaction strength $\tJ$ in defining the state of the 
combined system, rather than their actual individual values.

Non-GP solutions do exist for $|\tJ|>1$. When $|\tJ m^\sigma|>1$ we have
\begin{equation}
  \label{equation:mlowT}
  m^\sigma = \sgn\,J\cdot\sgn\,m^\sigma,
\end{equation}
and $m^\sigma$ becomes independent of $m^\tau$. Ignoring the GP solution, we are left with
$m^\sigma=\pm \sgn \, m^\sigma$ depending on the sign of the intraplane interaction.

\begin{figure}
 \center
 \includegraphics[width=10cm]{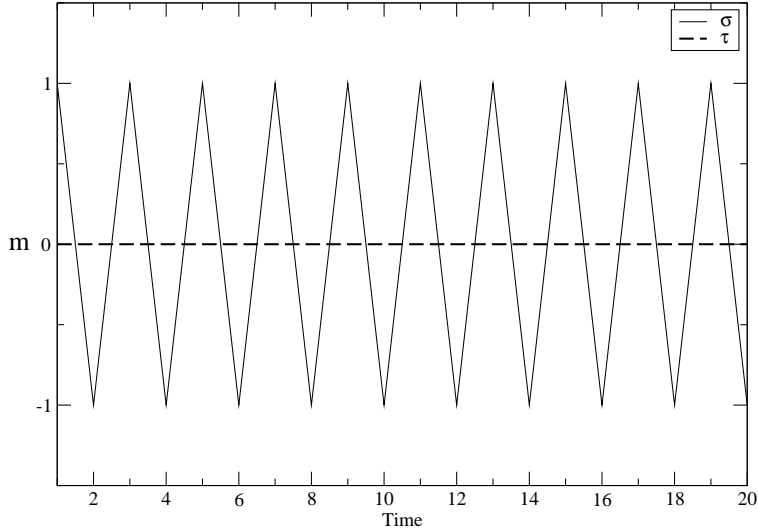}
 \caption{Dynamics of the 2D map of the magnetisations for $\beta_\tau=10^{-2}$, $\beta_\sigma=10^2$, $J=-5$ and $J'=10^{-2}$.
          The period 2 cycle in $m_\sigma$ sets in already at the first iteration, while $m_\tau$ remains at zero mainly due to
          its very high temperature.}
 \label{figure:P2C}
\end{figure}

It is convenient to consider several different cases:
(i) A positive sign gives $m^\sigma=\pm 1$, meaning that the intraplane interaction is sufficient to align the $\sigma$ component spins 
completely, independently of the heat flow from the $\tau$ to the $\sigma$-components. (ii) A negative sign results in oscillations
between $\pm1$, a cycle of period 2 for $m^\sigma$ instead of a fixed point
as depicted in Fig \ref{figure:P2C}.
This occurs, for instance, in the case antiferromagnetic intraplane interactions and is
characteristic of synchronous update dynamics.
(iii) Finally, the case $\tJ m^\sigma=\pm1$ has as the non-GP solutions
\begin{equation}
  m^\sigma = \pm\,\frac{\sgn\,J'}{2-\tbt},
\end{equation}
which exist when the interactions obey the \emph{temperature-dependent} condition $\tJ = (2-\tbt)\, \sgn J'$. This
means that for each $\tbt$ value \emph{only one} solution exists on the $\tJ$ line.

As the equations are completely symmetric with respect to exchanges of $\sigma$ and $\tau$, the only remaining case is the one where both
systems are at low temperatures with $T_\tau,T_\sigma\ll1$, or equivalently $|\tbt|,|\tbs|\gg1$. Once more we consider separately cases with different magnitudes of the interactions:

(a) \underline{\emph{For $|\tJ|<1$}} we have
\begin{align}
  f^\sigma &= m^\tau \sgn \, J',\\
  f^\tau   &= m^\sigma \sgn \, J'.
\end{align}
In this limit, all points of the $(m^\sigma,m^\tau)$-plane are either period-2 cycles or fixed points; the fixed points correspond to the line $m^\sigma=m^\tau \sgn \, J'$.

(b) \underline{\emph{When $|\tJ|=1$}}, the same solution is obtained for $|m^\sigma|,|m^\tau|<1$, but when $|m^\sigma|=|m^\tau|=1$ the fixed points
$m^\sigma=m^\tau \sgn \, J'$ occur only for $\sgn \, J'=\tJ$; otherwise the solutions are period-2 cycles.

(c) \underline{\emph{The $|\tJ|>1$ case}} is more complicated to analyse. Here the intraplane interaction is relatively strong and many different solutions can be found. We have to take into consideration one of the three cases for each of the
magnetisations ($a,b=\sigma,\tau$):
\begin{equation}
 \label{equation:cases}
 \begin{split}
   |\tJ m^a|>1 &\Rightarrow f^a = \sgn\,J \cdot \sgn\,m^a,\\
   |\tJ m^a|=1 &\Rightarrow f^a = \frac{m^b}{2\sgn\,J'-\tJ},\\
   |\tJ m^a|<1 &\Rightarrow f^a = m^b \sgn\,J'.
 \end{split}
\end{equation}

(i) The case of $|\tJ m^\sigma| >1$ and $|\tJ m^\tau|>1$  results in two independent equations of the form (\ref{equation:mlowT}) for each 
one of the magnetisations, giving completely aligned states or period-2 cycles for
all points with $|m^\sigma|>1/|\tJ|$ and $|m^\tau|>1/|\tJ|$. (ii) Another simple case is when $|\tJ m^\sigma|<1$ and $|\tJ m^\tau|<1$, or 
equivalently $|m^\sigma|<1/|\tJ|$ and $|m^\tau|<1/|\tJ|$, which reduces to the same set of solutions as for $|\tJ|<1$.

(iii) When $|\tJ m^\sigma|=|\tJ m^\tau|=1$, or $|m^\sigma|=|m^\tau|=1/|\tJ|$, the equations become
\begin{equation}
  m^\sigma = \frac{m^\tau}{2\sgn\,J'-\tJ}, \qquad  m^\tau = \frac{m^\sigma}{2\sgn\,J'-\tJ},
\end{equation}
which exists when the interactions obey the relation $2\sgn\,J'-\tJ=\pm1$. For the plus sign, we have fixed points
where $m^\sigma=m^\tau$, while the minus sign gives $m^\sigma=-m^\tau$.

It remains to analyse the mixed cases with: (iv) $|\tJ m^a|>1$, $|\tJ m^b|<1$; (v) $|\tJ m^a|>1$, $|\tJ m^b|=1$;
(vi) $|\tJ m^a|<1$, $|\tJ m^b|=1$.

Cases (iv) and (v) can only have non-GP phases for $m^a=\pm1$ and $\sgn\,J=+1$ and a 2-period oscillatory behaviour when $\sgn\,J=-1$ with 
$m^a$ oscillating between +1 and $-1$. In both cases, $m^b$ will be a  multiple of $m^a$ according to the
corresponding value given by one of the equations (\ref{equation:cases}). Finally, all points in case (vi) are either
fixed points, when $2-\tJ \sgn\,J'=+1$, or 2-period cycles, when $2-\tJ \sgn\,J'=-1$

\begin{figure}
  \centering
  \includegraphics[width=10cm]{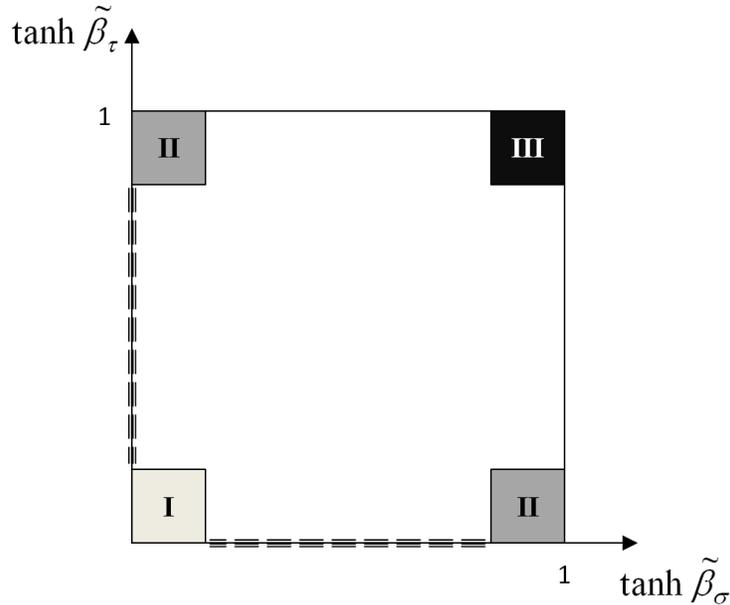}
  \caption{Partial phase diagram for the limiting cases of high and low temperatures. The figure shows only the first quadrant
           as the other three are simply its symmetrical images reflected about the axes. On the dashed axis, the
           corresponding magnetisation is zero as the temperature is infinite.}
  \label{figure:PD}
\end{figure}

Figure~\ref{figure:PD}, together with table~\ref{table:PD}, presents a summary of the possible asymptotic phases of the system in the
limits we analysed. Intermediate regimes are more difficult to analyse, but the equations are simple enough to be solved numerically for any 
parameter values.

\begin{table}
  \centering
  \caption{Phases of Fig \ref{figure:PD}.}
  \begin{tabular}{|c|l|}
    \hline
    \textbf{Phase} & \textbf{Description} \\
    \hline
    $\begin{array}{c}
      \text{I} \\ T_\sigma,T_\tau\gg1
    \end{array}$ & Global Paramagnetic (GP) only ($m^\sigma=m^\tau=0$) \\
    \hline
    $\begin{array}{c}
      \text{II} \\ T_a\gg1 \\ T_b\ll1 \\ a,b = \sigma,\tau
     \end{array}$ &
      $\begin{array}{cl}
        m^a \approx \tilde{\beta}_a m^b, \\
        |\tJ|\leq1  & \Rightarrow \text{GP-only}, \\
        |\tJ|>1     & \Rightarrow
                        \left\{\begin{array}{cl}
                          |\tJ m^b|<1 & \Rightarrow \text{GP-only}\\
                          \tJ m^b =\pm1 & \Rightarrow m^b=\pm \sgn\,J'/(2-\tilde{\beta}_a)\\
                          |\tJ m^b|>1 & \Rightarrow
                                        \left\{\begin{array}{cl}
                                          J>0 & \Rightarrow |m^b|=1 \\
                                          J<0 & \Rightarrow \text{Period-2 Cycle in } m^b,
                                        \end{array}\right.\\
                        \end{array}\right.\\
      \end{array}$\\
    \hline
    $\begin{array}{c}
     \text{III} \\ T_\sigma,T_\tau\ll1
    \end{array}$&
      $\begin{array}{cl}
        |\tJ|<1  & \Rightarrow \left\{\begin{array}{cl}
                        m^\sigma=m^\tau\sgn\,J' & \Rightarrow \text{Fixed Points}\\
                        \text{all other points} & \Rightarrow \text{Period-2 Cycles},\\
                      \end{array}\right.\\
        |\tJ|=1  &  \Rightarrow \left\{\begin{array}{cl}
                        |m^\sigma|,|m^\tau|<1 & \Rightarrow m^\sigma=m^\tau\sgn\,J'\\
                        |m^\sigma|=|m^\tau|=1 & \Rightarrow \left\{\begin{array}{cl}
                                                              \tJ=\sgn\,J' & \Rightarrow m^\sigma=m^\tau\sgn\,J',\\
                                                              \tJ=-\sgn\,J' & \Rightarrow \text{Period-2 Cycles},\\
                                                     \end{array}\right.
                      \end{array}\right.\\
        |\tJ|>1     & \Rightarrow \left\{\begin{array}{cl}
                        |\tJ m^\sigma|,|\tJ m^\sigma|>1 & \Rightarrow \left\{\begin{array}{cl}
                                                              J>0 & \Rightarrow \text{Totally Aligned},\\
                                                              J<0 & \Rightarrow \text{Period-2 Cycles},\\
                                                     \end{array}\right.\\
                        |\tJ m^\sigma|,|\tJ m^\sigma|<1 & \Rightarrow \text{Same as } |\tJ|<1,\\
                        |\tJ m^\sigma|,|\tJ m^\sigma|=1 & \Rightarrow \left\{\begin{array}{cl}
                                                              \Delta\equiv 2\,\sgn\,J'-\tJ=\pm1 & \Rightarrow
                                                                                                  m^\sigma=\Delta m^\tau,\\
                                                              \text{All other cases} & \Rightarrow \text{GP-only},\\
                                                     \end{array}\right.\\
                        |\tJ m^a|>1,|\tJ m^b|<1 & \Rightarrow \left\{\begin{array}{cl}
                                                              J>0 & \Rightarrow m^a=\pm1, m^b=m^a\,\sgn\,J',\\
                                                              J<0 & \Rightarrow \text{Period-2 Cycles},\\
                                                     \end{array}\right.\\
                        |\tJ m^a|>1,|\tJ m^b|=1 & \Rightarrow \left\{\begin{array}{cl}
                                                              J>0 & \Rightarrow m^a=\pm1, m^b=m^a/(2\,\sgn\,J'-\tJ),\\
                                                              J<0 & \Rightarrow \text{Period-2 Cycles},\\
                                                     \end{array}\right.\\                                                                               |\tJ m^a|<1,|\tJ m^b|=1 & \Rightarrow \left\{\begin{array}{cl}
                                                              2-\tJ\,\sgn J'=+1 & \Rightarrow \text{All fixed points},\\
                                                              2-\tJ\,\sgn J'=-1 & \Rightarrow \text{All period-2 cycles}.\\
                                                     \end{array}\right.\\
                      \end{array}\right.\\
      \end{array}$\\
    \hline
  \end{tabular}
  \label{table:PD}
\end{table}

Due to the non-linearity of the equations, we carried out extensive simulations with many different sets of parameters looking for an 
indication of chaotic behaviour, but did not find any. Presumably this is due to the fact that the equations of macroscopic order parameters 
result from a stochastic process fluctuating very closely to the typical values at the large system limit; as such the system is not amenable 
to chaotic behaviour.

\section{Stability of the GP Phase}
\label{section:PP}

As previously noted, there are two different types of paramagnetic phases in the presented model:
(i) the GP phase where $m^\sigma=m^\tau=0$, and the (ii) $a$-component paramagnetic phases, where $m^a=0$,
with $a$ either $\sigma$ or $\tau$ and the magnetisation of the other component being non-zero.

The GP phase is ubiquitous in the studied model, being always a solution for any combination of parameters.
However, depending on the values of these parameters, the stability of this phase can drastically vary; it
can be analysed by calculating the eigenvalues of the Jacobian matrix of our two-dimensional map. This general form of the matrix is 
\begin{equation}
  \begin{split}
    Df &= \left(
            \begin{array}{cc}
              \partial f^\sigma/\partial m^\sigma & \partial f^\sigma/\partial m^\tau \\
              \partial f^\tau/\partial m^\sigma & \partial f^\tau/\partial m^\tau
            \end{array}
          \right)\\
        &= \left(
            \begin{array}{cc}
              \bes J \avg{\sech^2 \tbs\prs{\tJ m^\sigma+\tau}}{\tau} & \frac12 \sum_{s=\pm1} \tanh \tbs(\tJ m^\sigma s+1) \\
              \frac12 \sum_{s=\pm1} \tanh \tbt(\tJ m^\tau s+1) & \bet J \avg{\sech^2 \tbt\prs{\tJ m^\tau+\sigma}}{\sigma}
            \end{array}
          \right).
  \end{split}
  \label{equation:Jacobian}
\end{equation}

For the GP phase, at $m^\sigma=m^\tau=0$, the Jacobian matrix simplifies to
\begin{equation}
  Df = \left(
          \begin{array}{cc}
            \bes J \sech^2 \tbs & \tanh \tbs \\
            \tanh \tbt          & \bet J \sech^2 \tbt
          \end{array}
        \right),
\end{equation}
with eigenvalues
\begin{equation}
  \lambda^{\pm} =\frac12 (-b\pm\sqrt{\Delta}),
\end{equation}
where
\begin{align}
  b      &= -J\prs{\bes\sech^2\tbs+\bet\sech^2\tbt},\\
  \Delta &= J^2\prs{\bes\sech^2\tbs-\bet\sech^2\tbt}^2+4\tanh\tbs\tanh\tbt.
\end{align}

\begin{figure}
 \center
 \includegraphics[width=7cm]{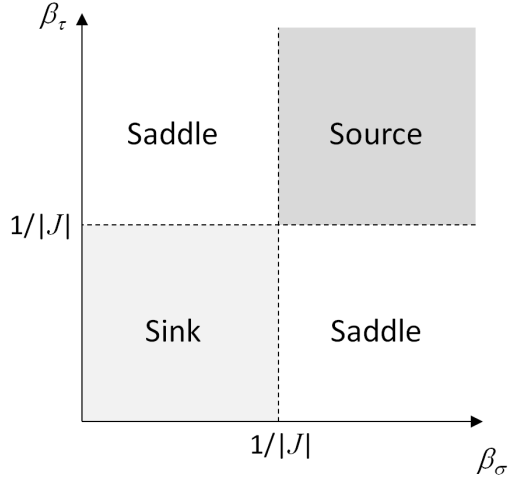}
 \caption{Stability of the GP phase for high temperatures (small $\beta$'s).}
 \label{figure:BHT}
\end{figure}

For $\tbt=\tbs=0$, both eigenvalues are zero indicating that the GP phase is a sink, i.e., stable to perturbations. For small,
but non-zero inverse temperatures (high-temperature limit), we have
\begin{equation}
  \lambda^{\pm} \approx \frac12 \col{J\prs{\beta_\sigma+\beta_\tau}\pm|J\prs{\beta_\sigma-\beta_\tau}|},
\end{equation}
which gives eigenvalues $J\beta_\sigma$ and $J\beta_\tau$. As stability in the direction of
each of the eigenvectors depends on the magnitude of the corresponding eigenvalue, we see that the GP phase becomes
unstable below some temperature value. For either $\beta_\sigma\gtrsim1/|J|$ or $\beta_\tau\gtrsim1/|J|$ separately, it becomes a 
\emph{saddle}. If both conditions are satisfied it becomes a \emph{source}. Note that these conditions are exactly the ones defining the 
critical point of the naive mean field Ising model. The results for high temperatures are indicated in Fig \ref{figure:BHT}.

The results for low temperatures with $\beta_\sigma,\beta_\tau\rightarrow\infty$ are
\begin{equation}
  \lambda^\pm \approx \pm1+2 J(\beta_\sigma e^{-2\tbs}+\beta_\tau e^{-2\tbt}),
\end{equation}
and the GP is always a saddle.

An interesting case occurs in the regime where one of the temperatures is high and the other is low.
By taking $\beta_\tau\rightarrow0$ and $\beta_\sigma\rightarrow\infty$, straightforward calculations show that
\begin{equation}
  \lambda^\pm \approx J\prs{2\beta_\sigma e^{-2\beta_\sigma}+\beta_\tau}\pm\sqrt{|J'|\beta_\tau}.
\end{equation}

\begin{figure}
  \centering
  \includegraphics[width=7cm]{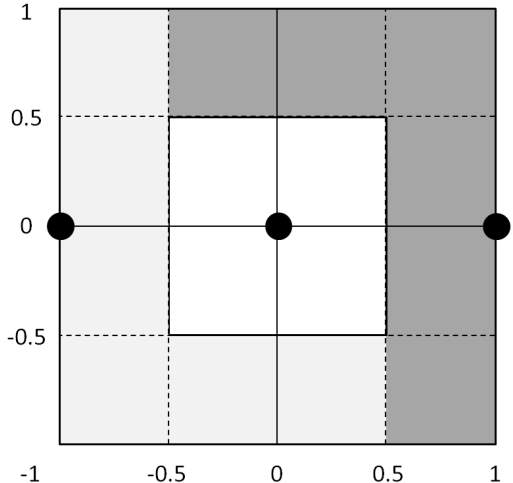}
  \caption{Basin of attraction for values $J=2$, $J'=1$, $\beta_\sigma=10^3$ and $\beta_\tau=0$.}
  \label{figure:BAGP}
\end{figure}

For sufficiently large values this points to a stability of the GP phase for all interaction values.
This can be seen from the basin of attraction depicted in Fig.~\ref{figure:BAGP}. The figure shows the
basin of attraction for fixed points of the two-dimensional map (\ref{eq:fs_ms}) with parameters given by $J=2$, $J'=1$,
$\beta_\sigma=10^3$ and $\beta_\tau=0$. The three black circles correspond to the stable fixed points and the shaded
regions around each of them are their respective basins of attraction.
There are also two unstable fixed points, located in the same horizontal line as the stable ones at the border of
the touching basins of attraction.


\section{Conclusions}
\label{section:Conc}

With the aim of understanding the behaviour of non-equilibrium systems with two-temperatures, we introduced and solved a
simple model of a system comprising two spin-1/2 components, referred to as the $\sigma$ and $\tau$ components; these are coupled to separate 
thermal reservoirs held at different temperatures $T_\sigma$ and $T_\tau$, respectively. For visualisation purposes, we considered the two 
system components as belonging to two different planes. Each
component has then an \emph{intraplane} long-ranged mean-field interaction $J$ while it interacts with the other component via an
\emph{interplane} local interaction $J'$.

Using the generating functional technique we have found a pair of exact, self-consistent coupled equations for the magnetisations of the two 
components $m^\sigma$ and $m^\tau$, the solution of which is given by the fixed-points of the resulting two-dimensional non-linear
map. This map depends on three degrees of freedom which are given by the ratio between the inter and intra-component couplings $\tJ=J/J'$ 
and the scaled inverse temperatures of each component $\tbs=J'\beta_\sigma$ and $\tbt=J'\beta_\tau$.

The equations for this model are reminiscent of equilibrium equations for mean-field magnetic systems and allow for numerical
solution of the fixed-points at any parameter values. Analytically, we have been able to
study the possible phases of this system in the limits of low and high temperature, where one could identify paramagnetic and
ferromagnetic phases, showing that these are well defined steady states.

The analysis of the particular phase where both components have zero magnetisation, which we termed the Global Paramagnetic
(GP) phase, show that the stability of this phase has some interesting characteristics, especially in the low and high
temperature limits. We have not found any evidence for chaotic behaviour in general, but have no definitive proof either for its existence 
or absence as each orbit would need to be analysed individually. We conjecture that this is due to the nature of the stochastic dynamics 
that give rise to this mapping as the magnetisations fluctuate very close to the mean values at the large system limit which we investigate.

The framework introduced and the results obtained can be easily modified to accommodate other cases; one
obvious modification being the spatial nature (local/non-local) of the interactions. Also, with the appropriate changes,
one can apply this theory to several other non-equilibrium practical situations where a steady state is reached.

One of these cases, which we are currently exploring, is related to astrophysical dark matter as the dark sector might have, in additional 
to its gravitational coupling to the visible sector, exclusive dark matter interactions. This opens up the possibility of systems at different 
temperatures to co-exist at the same physical location but in different sectors. These systems can then be heated independently by processes 
on their own sectors and exchange
energy only via gravitational interactions. The current framework could be modified to accommodate long-range gravitational ``interplane'' 
interaction  and short-rang ``intraplane'' interaction according to bounds obtained from
astrophysical observations~\cite{Gradwohl92}.  Analysis of the phases and the heat-transfer between both
sectors could shed light on the nature of these potential interactions or serve as an additional detecting mechanism.

Another extension of this framework, which is also under way, is the application to superlattices \cite{Tsu05}. 
Superlattices are metamaterials constructed 
of alternate layers of two different materials that hold useful properties. In a forthcoming study, we appropriately
alternate thermal conducting magnetic layers with thermal isolating ones such that each one of $n>2$ conducting layers can
be coupled to a different thermal reservoir. This would result in an $n$-temperatures systems and the magnetisations would
then be described by an $n$-dimensional non-linear map with the promise of a rich and interesting phase diagram.

We believe that there are potentially many cases of non-equilibrium physical systems comprising a number of components, each of which is held 
at a different temperature and exhibits interactions of a different nature to the cross-component ones. This analysis of the particularly 
simple case of interacting magnetic systems held at different temperatures, can be carried out rigorously and paves the way to better 
understanding of similar cases in a variety of fields.

\section*{Acknowledgments}

Support by the Leverhulme trust (F/00 250/M) is acknowledged. A.C. would like to acknowledge Dr Banibrata Mukhopadhyay of the Indian 
Institute of Science, Bangalore, for useful discussions.


\end{document}